%% file: openwork.tex
\documentclass[12pt]{article}
\usepackage{openwork}
\usepackage{virgil_custom}
\graphicspath{ {.} {./figs/} }

\title{Quantifying Spacetime Integration across a Partition with Synergy}
\author{Virgil Griffith \texttt{<arxiv@virgil.gr>}}
\affil{Georgetown University}
\date{}

\begin{document}

\maketitle

\begin{abstract}
In service to the mathematical underpinnings of the Information Integration Theory of Consciousness (IIT), we introduce four principled measures of integration based on the partial information decomposition framework. We compare our measures to current IIT practice in simple deterministic networks. Our measures are the first IIT-related state-dependent integration measures to naturally obey standard infotheoretic lower and upper bounds. Matching IIT4, the four measures are non-Shannon, but outside IIT they work just as well as standard Shannon-based measures of irreducibility in discrete dynamical systems. 
\end{abstract}

\section{Introduction}
Consciousness is a hard problem. The Information Integration Theory of Consciousness (IIT) is an evolving theory \cite{iit0, iit1, iit2, iit3, iit4} to tackle it and will continue evolving for some time. The most recent version, IIT4, has six aspects: Existence, Intrinsicality, Information, Integration, Exclusion, and Composition. Every aspect has undergone revision, and here we focus solely on the notion of \emph{integration}. There exist several preexisting, plausible, principled, mathematical definitions of integration \cite{mediano2021, phipid, langer2020} and we propose a measure of integration based on the reasonably well-established concept of \emph{synergistic information} \cite{williams2010, qsmi,e19090456}.

Synergy has been designed and rigorously vetted to quantify the magnitude of multivariate interactions and is often used to measure ``complexity'' or ``irreducibility to constituent information sources''. As IIT often frames integration as a sort of irreducibility, there seems to be nonnegligible overlap between IIT's integration and synergistic irreducibility. We see no major conflicts between synergy and IIT axioms/postulates, and synergy permits greater nuance and specificity in clarifying what IIT means by ``integration.''

In this paper, we: explain how IIT4 operationalizes integration across a partition; review synergy; show how synergy differs from IIT between 2008--2025; introduce \emph{four} not-unreasonable measures for spacetime integration across a partition derived from synergy; compare the first and third measures to IIT in simple deterministic systems; and conclude with recommendations for IIT's future treatment of integration. The first measure matches IIT4 as closely as possible. The other three aim to treat space and time in a more holistic, integrated manner.

\section{How Integration is Currently Done}
We define terms. IIT4 \cite{iit4} starts with a system random variable (r.v.) $S$ of $n$ nodes in a single state $s$ evolving by transition probability matrix $\mathcal{T}_S$. In the spirit that a system must ``take a difference and make a difference'', it looks both backwards in time (taking a difference) and forwards in time (making a difference). Tweaking \cite{iit4}'s notation, we define a random variable representing the system's past, $A$ (``Apriori''), assumed to be a discrete uniform distribution. We also define a random variable for the system's future, $Z$ (``Zukunft''). The causal structure is $A \overset{\mathcal{T}_S}{\to} S \overset{\mathcal{T}_S}{\to} Z$.

1. IIT4 first uses the ``intrinsic information''\footnote{This has been renamed to ``intrinsic specification'' in \cite{new_intrinsicdifference}.} measure from \cite{intrinsicinfo} to extract two states from $A$ and $Z$, called the ``maximal cause-effect state,'' extracting the state from $A$ that conveys the most intrinsic information about $s$ as well as the state from $Z$ that conveys the most intrinsic information about $s$. This is defined as,

\begin{align}
a^\prime &\equiv \underset{a \in A}{\arg\max} \ ii_c(s,a) = \underset{a \in A}{\arg\max} \ \  p_c(a|s) \log \frac{p_c(s|a)}{p_c(s)} \\
z^\prime &\equiv \underset{z \in Z}{\arg\max} \ ii_e(s,z) = \underset{z \in Z}{\arg\max} \ \ p_e(z|s) \log \frac{p_e(z|s)}{p_e(z)} \; .
\end{align}

The function $p_c(\cdot|\cdot)$ is defined by assuming the given state, $a^\prime$ or $s$, follows a discrete uniform distribution. The function $p_e(\cdot|s)$ is defined by assuming state $s$ and $p(s)=\sum_{a \in A} p(a,s)$. Collectively, the two transitions, $a^\prime \rightarrow s$ and $s \rightarrow z^\prime$, are the center of analysis.

2. After defining the intrinsic informations from $a^\prime \rightarrow s$ and $s \rightarrow z^\prime$, they ask how much of each is irreducible to disjoint parts. They define $\phi_c$ and $\phi_e$ that compute how much of each intrinsic information is irreducible to a given directional partition $\theta$ of $k$ parts where $2 \leq k \leq n$. It is defined by \cite{iit4}'s eqs.~(18) and (19),

\begin{eqnarray}
    \phi_c\left( \mathcal{T}_S, s, \theta \right) &\equiv& \left| p_c(a^\prime|s) \log \left( \frac{p_c(s|a^\prime)}{p_c^\theta(s|a^\prime)} \right) \right|_+\; \\
    \label{eq:phi_c}
    \phi_e\left( \mathcal{T}_S, s, \theta \right) &\equiv& \left| p_e(z^\prime|s) \log \left( \frac{p_e(z^\prime|s)}{p_e^\theta(z^\prime|s)} \right) \right|_+\; .
\end{eqnarray}

The function $p_e^\theta(\cdot|s)$ does the work in ``causal marginalizing'' along the cuts of partition $\theta$. These cuts are applied across all three states: $a^\prime$, $s$, and $z^\prime$. The $\phi_c$ and $\phi_e$ measures have two caveats. First without the $|\cdot|_+$ operation, they can be negative. The $|\cdot|_+$ sets any negative values to zero. Second, $\phi_c(s)$ and $\phi_e(s)$ can both exceed their respective intrinsic informations, ie, $\phi_c(s) \not\leq ii_c(a^\prime, s)$ and $\phi_e(s) \not\leq ii_e(s,z^\prime)$.\footnote{In future work (eq.23 in \cite{new_intrinsicdifference}), this is accounted for by setting the $\phi$ value exceeding the intrinsic information to the intrinsic information.} To better understand $p_c(\cdot|\cdot)$ and $p_e(\cdot|\cdot)$ please see \cite{iit4}.

3. To capture the conjunction of ``taking and making a difference'', the final irreducible intrinsic information measure of a state $s$ over a partition $\theta$ is $\phi_s$, the minimum\footnote{Technical point: IIT4 aims to measure the ``conjunction'' across time of spatial integrations, quoting \cite{iit4}'s Principle of Being: ``A similar principle can be found in the work of the Buddhist philosopher Dharmakīrti: `Whatever has causal powers, that really exists.' Note that the Eleatic principle is enunciated as a disjunction (either to do something... or to be affected...), whereas IIT’s principle of being is presented as a conjunction (take and make a difference).'' Even presuming we have the perfect measures of spatial integration $\phi_c$ and $\phi_e$, $\phi_s$ does not measure the conjunction across time. Within both probability and information theory, given $f(A)$ and $f(B)$, the conjunction $f_\cap(A, B) \leq \min [f(A), f(B)]$. For example, this holds for: probability density, entropy, and mutual information. So taking minimum merely \emph{upperbounds} the true conjunction of spatial integration. One way one could calculate the conjunction of the \emph{total} information, not solely the synergistic part, is simply the $\operatorname{I}_\cap\left(\{A,Z\} \to s\right)$ from the PID framework. At current mathematical understanding, there's no known way to get only the conjunctive synergy present from both directions. However, Spivak's category theory\cite{spivak2013categorytheoryscientistsold} would be a place to start.} of the cause and effect irreducible intrinsic information,
\begin{equation}
    \label{eq:phi_s_iit4}
    \phi_s( \mathcal{T}_S, s, \theta ) \equiv \min\left[\phi_c\left( \mathcal{T}_S, s, \theta \right), \phi_e\left( \mathcal{T}_S, s, \theta \right) \right] \; .
\end{equation}

4. To decide which partition $\theta$ is the ``fault line'' of the system in state $s$, \cite{iit4} defines the ``Minimum Information Partition'' (MIP). The MIP is the directional partition $\theta$ that yields the lowest normalized $\phi_s$, an algorithm of $O(n!)$.

5. Finally, the intrinsic integration for the state $s$ is the \emph{unnormalized} $\phi_s$ over state $s$'s MIP found in step four.

\section{Understanding Synergy}
Synergy is the nonnegative slice of the mutual information between a set of random variables and a single target random variable that is irreducible to any single r.v. from the set.\cite{williams2010, qsmi, Kolchinsky_2022} Given a set of $n$ random variables $\{X_1, \ldots, X_n\}$ as \emph{predictors} of the \emph{target} random variable $Y$, it is defined as the whole minus the ``union'' information between every predictor and the target.

The first step is to define the ``union'' probability distribution $p_\cup$,
\begin{equation}
\label{eq:pcup}
p_\cup\left( X_1, \ldots, X_n, Y\right) \equiv \underset{p(\hat{X}_1,\ldots,\hat{X}_n,Y)}{\operatorname{argmin} } \operatorname{I}\left( \hat{X}_1, \ldots, \hat{X}_n ; Y \right) \ \ \ \textnormal{s.t. $p(\hat{X}_i,Y)=p(X_i,Y)\ \forall i$} \; .
\end{equation}

Algorithms for the minimization in eq.~\eqref{eq:pcup} exist for arbitrary $n$.\cite{syn_arbitraryN, synergy_arbitrary} Very efficient algorithms exist for $n=2$ \cite{makkeh2018broja} and $n=3$ \cite{nequals3}.

Once we have the $p_\cup$ distribution, computing the synergy, $\mathcal{S}$, is straight-forward:

\begin{eqnarray}
\label{eq:S}
    \mathcal{S}\left( \{ X_1, \ldots, X_n \} \rightarrow Y\right) &\equiv& \Ip{X_1, \ldots, X_n}{Y} - I_{\cup}\left( X_1, \ldots, X_n ; Y \right) \\
    &=& \Ip{X_1, \ldots, X_n}{Y} - \operatorname{D_{KL}}\left[ p_\cup(X_1,\ldots,X_n,Y) \mid\mid p_\cup(X_1,\ldots, X_n) p_\cup(Y) \right] \nonumber \; .
\end{eqnarray}

Synergy has some known bounds, with the most general from \cite{Kolchinsky_2022},
\begin{equation}
    \label{eq:mipbound}
    \mathcal{S}(\{X_1,\ldots, X_n\} \rightarrow Y) \leq \min_i \Ip{X_1, \ldots, X_n}{Y|X_i} \; .
\end{equation}

For all prior synergy work, the measure is averaged over all states of target $Y$ and uses the $\operatorname{D_{KL}}$. However, since IIT4, IIT prefers the Intrinsic Difference\footnote{For those circumspect of non-Shannon measures, $\operatorname{ID}$ is the same thing as $\operatorname{D_{KL}}$ but replaces the $\sum_{x_1,\ldots,x_n}$ with $\max_{x_1,\ldots,x_n}$. Let your fears be assuaged.} (ID) \cite{intrinsicinfo} into a specific state $y \in Y$, defined as,

\begin{equation}
    \operatorname{ID}\left[ p(X_1, \ldots, X_n) \to y \right] \equiv \max_{x_1,\ldots,x_n} p(x_1, \ldots, x_n|y)  \log \frac{p(x_1, \ldots, x_n|y)}{p(x_1, \ldots, x_n)} \; .
\end{equation}

We are not arguing that all those using Shannon-based partial information decomposition synergy should instead use the Intrinsic Difference. All measures in the following sections can be done with standard Shannon measures. For those preferring to stay within Shannon theory, we recommend doing exactly that. We merely show that synergy is flexible enough to \emph{also} be applied to IIT's Intrinsic Difference. Calculating synergy for an intrinsic difference requires only modest adaptation.\footnote{Specifically, we use the same $p_\cup$ distribution. There are alternative ways of defining $\mathcal{S}_{ID}$, see Appendix \ref{sect:otherway}.} We define the \emph{synergistic intrinsic difference}, $\mathcal{S}_{ID}$, into a single state $y$ as,

\begin{equation}
    \label{eq:SID}
    \mathcal{S}_{ID}\left( \{X_1, \ldots, X_n\} \rightarrow y \right) \equiv  \max_{x_1,\ldots,x_n} p(x_1, \ldots, x_n|y)  \log \frac{p(x_1, \ldots, x_n|y)}{p(x_1, \ldots, x_n)} - p_\cup(x_1,\ldots,x_n|y) \log \frac{p_\cup(x_1,\ldots,x_n|y)}{p_\cup(x_1,\ldots,x_n)} .
\end{equation}

Just as in the Shannon-case, $\mathcal{S}_{ID}$ is nonnegative. It is nonnegative because eq.~\eqref{eq:S} is nonnegative, and eq.~\eqref{eq:SID} is simply the largest ``nonnegative slice'' of eq.~\eqref{eq:S}. Unfortunately, getting upperbounds for $\mathcal{S}_{ID}$ is not as simple eq.~\eqref{eq:mipbound} due to the lack of averaging. Until tighter bounds emerge, we can use the upperbound,

\begin{equation}
    \label{eq:mipboundid} 
\mathcal{S}_{ID}(\{X_1,\ldots, X_n\} \rightarrow y) \leq  \operatorname{ID}\left[p(X_1,\ldots,X_n) \to y\right] \; .
\end{equation}


\subsection{Main Benefits of Synergy}
\begin{itemize}[noitemsep]
    \item Our measures of synergy use the ``unitary present''. The partition is applied to the Apriori and Zukunft distributions, but not onto the current state $s$. We view this as more consistent with treating the entire system, in its current moment, as an undivided entity.\footnote{One could of course partition state $s$ according to $\theta$ and compute synergy going into each part of $s$, but we feel this doesn't treat the system as a whole and moreover introduces complications in double-counting synergies.}
    \item Although inspired by IIT postulates, maintaining the ``unitary present'' presents three promising theoretical simplifications: 
    
    \begin{itemize}
        \item \textbf{Naturally obeying bounds.} Current $\phi$ \cite{iit4,new_intrinsicdifference} is a bit kludgy in how it stays within its intuitive bounds of zero and the intrinsic information. It does this by taking any values outside of those bounds forcing them back into the admissible range. Synergy needs no kludges.
        \item \textbf{Less causal marginalization?} In IIT4 we causally marginalize when applying the partition $\theta$ in computing $p_e^\theta(\cdot|\cdot)$. By never partitioning the present state, that may no longer be needed. However we will still need to causally condition when finding Complexes.
        \item \textbf{Simplified partitions?} Our understanding is that directional partitions were added to account for information moving across parts in a single direction. Within a synergy framework, this is considered acceptable (eg, AND-ZERO in Table \ref{fig:one} in next section). Therefore directional partitions may no longer be needed.
    \end{itemize}
    
    \item Synergy measures, in a very principled way, a joint information's irreducibility to constituent informations. This seems appropriate for a theory that puts irreducible information at front and center.
    \item Synergy yields, in our view, a stronger correlation and connection between ``irreducibility'' and ``complexity''. This relationship may be nonessential, but it's certainly nice to have.

\end{itemize}

\section{Demonstrating the Promise of Synergy}

The synergy framework does not create a maximal cause-effect pair. Instead, it looks over the entire distribution of $A$ and $Z$ and applies the partition $\theta$ over $A$ yielding $\{A_1, \ldots, A_k\}$ and over $Z$ yielding $\{Z_1, \ldots, Z_k\}$. By design, we keep the current state $s \in S$ whole.

We now introduce our first measure of integration based on synergy and compare it to IIT4. In the next section we introduce three more ``elegant'' measures of spacetime integration that handle space and time more holistically.

\textbf{Measure 1: Matching IIT4.} The first of our four measures matches IIT4. It is based on $\phi_c$ \cite{iit4}'s eq.~(20),
\begin{equation}
    \label{eq:phi_c_S1}
    \phi_c^{S1} (s,\theta) \equiv \mathcal{S_{ID}}\left( \{A_1, \ldots, A_k\} \rightarrow s \right) \; .
\end{equation}

We do the same for $\phi_e$,

\begin{equation}
    \phi_e^{S1} (s,\theta) \equiv \mathcal{S_{ID}}\left( \{Z_1, \ldots, Z_k\} \rightarrow s \right) \; .
\end{equation}

And then to match the method in eq.~\eqref{eq:phi_s_iit4},

\begin{equation}
    \phi_s^{S1}(s,\theta) \equiv \min \left[ \phi_c^{S1} (s,\theta), \phi_e^{S1} (s,\theta) \right] \; .
\end{equation}


We wish to build intuition for how synergy captures integration differently from IIT4. We originally directly compared $\phi_s$ vs $\phi_s^{S1}$, but $\phi_s^{S1}$ was so strict having zero for all example systems and thus not very enlightening (results in Appendix \ref{app:comparefull}). So for pedagogical purposes of understanding ``what synergy is doing'' relative to IIT, we exploit symmetry and compare only the ``cause'' $\phi_c$ values across: IIT2 (2008), IIT3 (2014), IIT4 (2023), and the latest\cite{new_intrinsicdifference} (2025). After understanding $\phi_c$ then we can generalize to $\phi_s$ with greater confidence we are measuring what we intend. Of these, the most important comparison is between $\phi^{2023}_c$ (eq.~\eqref{eq:phi_c}) vs $\phi_c^{S1}$ (eq.~\eqref{eq:phi_c_S1}).

Table \ref{fig:one} is a battery of deterministic doublet systems. See Appendix \ref{app:tutorial} to understand these systems. We choose deterministic for simplicity and doublets to remove any ambiguity about the MIP. As all $\phi$ values vary with each state $s \in S$, for ease of comparison, we report the \emph{expected} $\phi_c$ over all states $s \in S$, denoted $\langle \phi_c \rangle$.\footnote{Examining the state-dependent case does not change the qualitative results for any presented examples.} 

Measure $\phi_c^{2025}$ is, by explicit design, zero for all deterministic systems regardless of their structure. While we are not opposed to this in principle, it is unclear to us why, on theoretical grounds, to preclude deterministic systems from consciousness. However, if this is a hard requirement for IIT measures, it can be accommodated within a synergy framework.\footnote{As we'll see with measure $\phi_s^{S4}$ in the next section.}

\begin{figure}[htb]
\centering

\begin{subfigure}{1.3in}
  \centering
  \includegraphics[width=\linewidth]{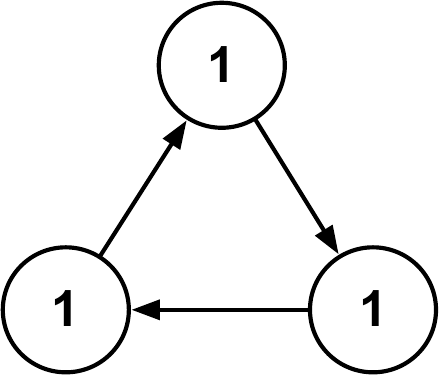}
  \caption{GET3}
  \label{fig:SHIFT}
\end{subfigure}
\hfill
\begin{subfigure}{1.3in}
  \centering
  \includegraphics[width=\linewidth]{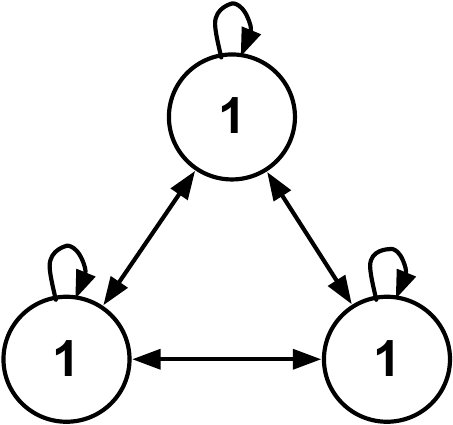}
  \caption{111}
  \label{fig:333}
\end{subfigure}
\hfill
\begin{subfigure}{1.3in}
  \centering
  \includegraphics[width=\linewidth]{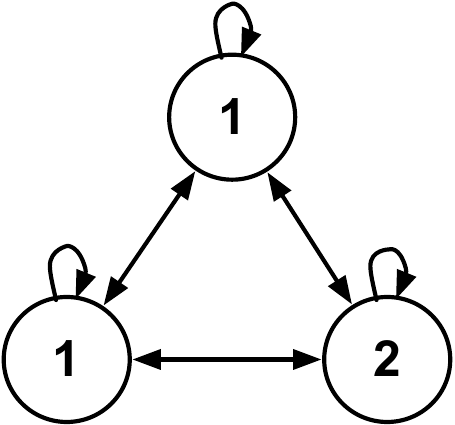}
  \caption{121}
  \label{fig:322}
\end{subfigure}
\hfill
\begin{subfigure}{1.3in}
  \centering
  \includegraphics[width=\linewidth]{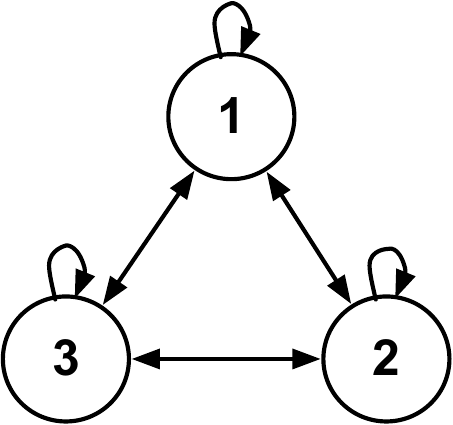}
  \caption{123}
  \label{fig:321}
\end{subfigure}

\vspace{0.3in}

\begin{tabular}{l | c c c c c | c c} \toprule
	Network & $\operatorname{ID}\left[ A \to S \right]$ & $\displaystyle \min_{s} \phi^{2023}_c(s)$ & $\displaystyle \max_{s} \phi^{2023}_c(s)$ & $\bphi$ & $\bphics$ & $\Ip{A}{S}$ & $\mathcal{S}_c$ \\
\midrule
	GET3      & 3.000 & 2.000 & 2.000 & 2.000 & 0 & 3.000 & 0 \\
	111       & 0.399 & 0 & 6.000 & 0.750 & 0.149 & 0.544 & 0.250 \\
	121       & 0.677 & 0 & 0.415 & 0.052 & 0.426 & 1.406 & 0.750 \\
	123       & 1.104 & 0 & 0 & 0 & 0.604 & 1.811 & 1.000 \\
\bottomrule
\end{tabular}
\caption{We compare IIT4's $\bphi$ to $\bpsi$ more carefully in triplet networks. We give the min and max $\phi_c^{2023}$ values to provide more insight into why $\phi_c^{2023}$ drops so precipitously. In our view, a more intuitive complexity measure would increase left to right, which $\bpsi$ does, yet $\langle \phi_c^{2023} \rangle$ does the opposite. To avoid claims of sleight-of-hand, $\phi_c$ is computed using a normalized MIP and $\phi_s^{S1}$ is computed using an unnormalized MIP. We are seeing $\phi^{2023}_c$ at its best, and $\phi_c^{S1}$ at its worst.}
\label{fig:diverse}
\end{figure}

Table \ref{fig:one} reveals three major differences between existing $\phi$ and synergy-based measures:
\begin{itemize}
    \item \textbf{Irreducibility: spatial or infotheoretic?} The clearest contrast between current IIT and synergy is in how they handle the GET (cross-copy) operation. Under spatial irreducibility measures $\phi^{2008}$, $\phi_c^{2014}$, and $\phi_c^{2023}$, GET operations are \emph{maximally} integrative. But under infotheoretic irreducibility measures like $\phi_c^{S1}$, GET operations are \emph{non}-integrative. Synergy requires at least two wires coming into the part, current IIT does not. The philosophical argument for why integration should be zero for GET-GET is that there's no place where the information ``comes together''. For example, if a brain was composed of two identical hemispheres each simply copying their states back and forth akin to GET-GET, does that whole brain possess a single conscious experience? We would argue ``No''.
    
    IIT is a theory based on irreducibility. But irreducible to what exactly? The mathematical argument favor of GET-GET's irreducibility is that each node is wholly determined by the other node; so GET-GET is \emph{spatially} irreducible. The mathematical argument against GET-GET's irreducibility is that it is \emph{infotheoretically} entirely reducible to the two information flows $\Ip{A_1}{S}$ and $\Ip{A_2}{S}$. There's nothing conceptually wrong with spatial irreducibility, however, infotheoretic irreducibility is mathematically deeper and arguably more conceptually appropriate for a theory as abstract as IIT.

    \item \textbf{What bounds?} Measures $\langle \phi^{2008} \rangle$, $\langle \mathcal{S}_c \rangle$, and $\langle \phi_c^{S1} \rangle$ are upperbounded by $\Ip{A}{S}$, $\Ip{A}{S}$, and $\operatorname{ID}[ A \to S]$ respectively. This is not the case with $\langle \phi_c^{2014} \rangle$ or $\langle \phi_c^{2023} \rangle$. See $\langle \phi_c^{2014} \rangle$'s AND-AND and $\langle \phi_c^{2023} \rangle$'s XOR-XOR as examples. Moreover, in the state-dependent case, $\phi^{2008}$, $\phi_c^{2014}$, and $\phi_c^{2023}$ can all exceed the entropy of the entire system. To an information theorist, this is akin to exceeding the speed of light. Shannon or otherwise, we feel it's very sensible for any information measure to be bounded by the corresponding entropy. Looking at $\langle \phi_c^{2014} \rangle$'s AND-GET, we see that sometimes \emph{even the average} case exceeds the total entropy (two bits). Violations of these upperbounds, in either the state-dependent or average case, imply these measures aren't quite capturing the intended quantity.

    \item \textbf{Is irreducibility consonant with complexity?} While irreducibility needn't be synonymous with ``complexity'', everything else equal, we'd prefer they be at least related. Complexity is usually thought of as combining ``interdependence'' and ``diversity''. Examples AND-OR and AND-XOR have more interdependence and diversity and thus are presumably more ``complex'' than GET-GET. Yet current IIT scores both AND-OR and AND-XOR as less integrated than GET-GET---with AND-OR's $\phi_c^{2023}$ even being zero. However, infotheoretic irreducibility measure $\phi_s^{S1}$ reasonably corresponds with complexity intuitions.
    
    Figure \ref{fig:diverse} shows that spatial irreducibility often has a inverse relationship with complexity. In Figure \ref{fig:diverse}, ``complexity'' should increase from left to right, which we do see in the infotheoretic irreducibility measures $\phi_c^{S1}$ and $\mathcal{S}_c$. Unfortunately, we see the opposite from $\phi_c^{2023}$. These are not cherry-picked examples; in Appendix \ref{appendix:4node} we see the same in four node networks. Until there’s a compelling argument that the spatial irreducibility present in GET3 is sufficient for awareness, from a complexity perspective it seems sensible to err on the side of infotheoretic irreducibility over spatial irreducibility.
\end{itemize}

\begin{table}[htb]
\centering
\begin{tblr}{l|c c c c c c| c c}
\toprule
\textbf{Doublet} & $\operatorname{ID}\left[ A \to S \right]$ & $\langle \phi^{2008} \rangle$ & $\langle \phi_c^{2014} \rangle$ & $\bphi$ & $\langle \phi_c^{2025} \rangle$ & $\bphics$ & $\Ip{A}{S}$ & $\mathcal{S}_c$ \\
\midrule
ZERO-ZERO & 0 & 0 & 0 & 0 & 0 & 0 & 0 & 0 \\
KEEP-ZERO & 0.500 & 0 & 0 & 0 & 0 & 0 & 1.000 & 0 \\
KEEP-KEEP & 2.000 & 0 & 0 & 0 & 0 & 0 & 2.000 & 0 \\
GET-ZERO & 0.500 & 1.000& 0 & 0 & 0 & 0 & 1.000 & 0 \\
GET-KEEP & 0.500 & 0 & 0 & 0 & 0 & 0 & 1.000 & 0 \\
GET-GET & 2.000 & 2.000 & 2.000 & 2.000 & 0 & 0 & 2.000 & 0 \\
\midrule
AND-ZERO & 0.604 & 0.500 & 0 & 0 & 0 & 0.354 & 0.811 & 0.500 \\
AND-KEEP & 1.250 & 0.189 & 0 & 0 & 0 & 0 & 1.500 & 0 \\
AND-GET  & 1.250 & 1.189 & 2.207 & 0.750 & 0 & 0 & 1.500 & 0 \\
AND-AND  & 0.604 & 0.189 & 1.484 & 0.500 & 0 & 0.354 & 0.811 & 0.500 \\
AND-OR   & 1.250 & 0.877 & 1.623 & 0     & 0 & 0.750 & 1.500 & 1.000 \\
AND-XOR  & 1.250 & 1.189 & 1.665 & 0.500 & 0 & 0.750 & 1.500 & 1.000 \\
\midrule
XOR-ZERO & 0.500 & 1.000 & 0 & 0 & 0 & 0.500 & 1.000 & 1.000 \\
XOR-KEEP & 2.000 & 1.000 & 0 & 0 & 0 & 0 & 2.000 & 0\\
XOR-GET & 2.000 & 2.000 & 1.500 & 2.000 & 0 & 0 & 2.000 & 0 \\
XOR-AND & 1.250 & 1.189 & 1.665 & 0.500 & 0 & 0.750 & 1.500 & 1.000 \\
XOR-OR  & 1.250 & 1.189 & 1.665 & 0.500 & 0 & 0.750 & 1.500 & 1.000 \\
XOR-XOR & 0.500 & 1.000 & 1.000 & 1.000 & 0 & 0.500 & 1.000 & 1.000 \\
\bottomrule
\end{tblr} 

\caption{Comparing $\phi^{2008}$, $\phi_c^{2014}$, $\phi_c^{2023}$, and $\phi_c^{S1}$ in simple networks. $\phi^{2008}$ only looked into the past and thus does not have a subscript $c$. For readers preferring Shannon-based measures, we include the mutual information $\Ip{A}{S}$ and synergistic information $\mathcal{S}_c(\{A_1,A_2\} \to S)$.}
\label{fig:one}
\end{table}

\section{Three Elegant Synergistic Measures of Integration}
\label{sect:elegant}
In Table \ref{fig:one}, we hope to have shown: The $\phi$ measure has evolved a great deal from 2008 to 2025, and that integration-as-synergy is a reasonable perspective. 

After adopting an integration-as-synergy perspective, some new options beyond the method that most closely matches IIT4 emerge. These new options allow us to be notably more nuanced and precise defining infotheoretic irreducibility across spacetime, specifically:

\begin{itemize}[noitemsep]
    \item Instead of simply taking the minimum between $\phi^{S1}_c$ and $\phi^{S1}_e$, we can directly quantify the extent to which the past and future \emph{irreducibly cooperate} in specifying the current state. This seems desirable for a measure of spacetime integration.
    \item IIT's aim to quantify the extent to which a system, ``takes a difference and makes a difference'' is often interpreted as desiring the T-symmetry that one repeatedly sees in physics. With the synergy framework, we can go beyond T-symmetry to make space and time \emph{fully interchangeable}. This seems appropriate for a theory that aims to define at what \emph{spatio-temporal grain} things exist. We aim to be of service.
    \item Modern phi aims to be \emph{zero} for deterministic systems.\cite{new_intrinsicdifference} We aim to be of service.
\end{itemize}

\textbf{Measure 2: Total Integrative Glue.} Looking forward and backwards, the total intrinsic information going into state $s$ is $\operatorname{ID}[ p(A, Z) \rightarrow s]$. Each partition $\theta$ breaks $A$ and $Z$ into $k$ parts each. We define the set $\Omega_\theta \equiv \{A_1, \ldots, A_k, Z_1, \ldots, Z_k\}$ as $\theta$'s ``sources'' of information about state $s$. The first question, how much of the total intrinsic information is irreducible to any single source? This is simply the synergy across all sources,

\begin{equation}
    \phi_s^{S2}(s,\theta) \equiv \mathcal{S}_{ID}\left( \Omega_\theta \rightarrow s \right) \; .
\end{equation}

The measure $\phi_s^{S2}$ quantifies the slice of the mutual information $\Ip{AZ}{s}$ that is both: intrinsic per intrinsic difference as well as irreducible to \emph{any single source} in the past or future---call it the ``total intrinsic spacetime integrative glue''. We can think of $\phi_s^{S2}$ as the total weight of the integrative tentacles reaching through spacetime to specify state $s$. To make an analogy, one can think of $\phi_s^{S2}$ as akin to the Binding Information from \cite{bindinginfo} but instead of quantifying the ``binding'' slices of a joint entropy, it is $\theta$'s ``intrinsic and binding'' slices of $\Ip{AZ}{s}$. Whereas $\phi_{s}^{S1}$ is the minimum irreducibility facing the past and future, $\phi_{s}^{S2}$ is the total irreducibility across past and future. One might think that $\phi_s^{S2}$ is ``disjunctive'' across time. It is not. A disjunctive integration across past and future would have $\phi_s^{S2} \geq \max \left[ \phi_c^{S1}, \phi_e^{S1} \right]$. However, if a part in the future specified information that was only synergistically specified in the past, then $\phi_s^{S2} < \phi_c^{S1}$. To be pedantic, $\phi_s^{S2} \lessgtr \phi_s^{S1}$.

One upside of $\phi_s^{S2}$ is that it is \emph{spacetime-interchangeable}, meaning that space and time are treated as fully equivalent. One downside is it might be felt as too loose. Specifically, some systems will have synergistic information \emph{solely across space} (eg, $\{A_1,A_2\} \to s$ or $\{Z_1,Z_2\} \to s$) or \emph{solely across time} (eg, $\{A_1,Z_1\} \to s$ or $\{A_2,Z_2\} \to s$), and $\phi_s^{S2}$ would consider that integrative.

\textbf{Measure 3: Disjoint parts through time.} Suppose we don't like the ST-interchangeability---after all, our conscious experiences of space \cite{tononispace} and time \cite{tononitime} are distinct. Moreover, within IIT we break time into two parts but space into $k$ parts, so space and time are already treated differently. What if we \emph{prefer} to treat space and time differently, but we still want T-symmetry?

This can be done. We take the $2k$ sources in $\Omega_\theta$ and merge them across time. This precludes any synergy across time within each part, ie, $\{A_i,Z_i\} \to s\ \forall i$. This measures the irreducibility beyond not $2k$ sources but $k$ \emph{spatially disjoint parts} across time. If one takes the intuition ``whole beyond the $k$ disjoint parts operating independently'' from \cite{iit2, langer2020}, and make it T-symmetric, it is a synergy over $k$ predictors where each predictor is a joint r.v. of a part in the past and future,

\begin{equation}
    \phi_s^{S3}(s,\theta) \equiv \mathcal{S}_{ID}\left( \{A_1 Z_1, \ldots, A_k Z_k\} \rightarrow s \right) \; .
\end{equation}

Measure $\phi_s^{S3}$ is practically computable today for bipartitions and tripartitions. In Appendix \ref{app:comparefull} we compare $\phi_s^{S3}$ to prior time-symmetric $\phi$-based measures across our simple examples. Measure $\phi_s^{S3}$ has well-behaved bounds and overall yields values more reasonably aligned with intuitions of ``complexity'' than $\phi_s^{2023}$, but it's not a perfect match. For example, $\phi_s^{S3}$ scores system 121 ($\langle \phi_s^{S3} \rangle=0.299$ bits) with higher integration than system 123 ($\langle \phi_s^{S3} \rangle=0.236$ bits). One interesting point is that $\phi_s^{S3}$ better matches complexity intuitions when using Intrinsic Difference than the Shannon variant $\mathcal{S}_{s}^{S3}$. This could be an argument for preferring the Intrinsic Difference over Shannon Information.

\textbf{Measure 4: Spacetime Unity.} Suppose we like the ST-interchangeability, but we want something stricter---the strictest even. We could count only that information that is irreducible to \emph{any subset} of sources across spacetime? This can be done. Like the more permissive $\phi_s^{S2}$, this strict variant is a synergy over $2k$ terms,

\begin{equation}
    \phi_s^{S4}(s,\theta) \equiv \mathcal{S}_{ID}\left( \{ \Omega_\theta \setminus x | x \in \Omega_\theta \} \rightarrow s\right) \; .
\end{equation}

This is the slice of $\Ip{AZ}{s}$ that is both intrinsic (per intrinsic difference) as well as irreducible to any subset of sources\footnote{Using the maximum over the subsets merely lowerbounds $\phi_s^{S4}$ as the max across subsets is less than or equal to the union across subsets.} drawn from the past or future. Think of it as quantifying how much the parts of $\theta$ constitute a unitary whole encompassing both the past and future. This measure is sufficiently strict that it seems that, regardless of chosen framework, $\phi_s^{S4}$ would lowerbound any plausible notion of state $s$'s spacetime integration. For deterministic systems, measure $\phi_s^{S4}$ is zero as some predictors contain the entirety of $A$.

\textbf{Relations.} For a given partition $\theta$ and state $s$, the three ``elegant'' measures obey the bounds,
\begin{equation}
    \label{eq:relations}
    0 \leq \phi_s^{S4} \leq \phi_s^{S3} \leq \phi_s^{S2} \leq \operatorname{ID}\left[ p(A,Z) \rightarrow s \right]\; .
\end{equation}

We summarize our four measures in Table \ref{tbl:summary}.

\begin{table}
\begin{tabular}{l c c c c c } \toprule
	 & \#terms & Symmetry? & Strictness & Irreducibility to $\theta$'s... & Intuition \\
\midrule
	$\phi_s^{S1}$      & $2 \times k$ & T-symmetric & medium & min past/future & matching IIT4 \\
    $\phi_s^{S2}$      & $2k$ & ST-interchangeable & low & ST-elements & total ST-integrative glue \\
	$\phi_s^{S3}$       & $k$ & T-symmetric & medium & spatial parts  & disjoint parts across time  \\
    $\phi_s^{S4}$       & $2k$ & ST-interchangeable & high & ST-subsets  & ST-unitary irreducibility \\
\bottomrule
\end{tabular}

\caption{Summarizing our four measures. Although eq.~\eqref{eq:relations} unifies three of our four measures, there's no simple relation among the rest. Specifically: $\phi_s^{2023} \lessgtr \phi_s^{S1} \lessgtr \phi_s^{S2}$.}
\label{tbl:summary}
\end{table}

\section{\emph{Minimum} or \emph{Modular} Information Partition?}
\label{mipbounds}
When using any synergy-based measures, the inequality in eq.~\eqref{eq:mipboundid} can be used for the normalization when finding the MIP, or even continue using the current MIP normalized from IIT4. So if finding the MIP via normalizing the $\phi_s$ is preferred, we can keep doing that. No change.

That said, is normalized $\phi_s$ the best way to find the MIP? If the goal is to exactly measure how irreducible the system in state $s$ is to \emph{any possible partition}, the set of all partitions includes the lop-sided partitions of just shaving off one node, so no normalization. And if there's no normalization, per \cite{psipaper1}'s Appendix B.1, a synergy measure's MIP will always be a bipartition thus reducing the MIP-finding algorithm from $O(n!)$ to $O(2^n)$.

However, if instead the goal is to measure how irreducible the system in state $s$ is to the state's ``fault lines''\footnote{Quoting from \cite{iit4}, ``the fault line dividing a system into two large subsets of units linked through a few interconnected units (a `bridge'), rather than defaulting to partitions between individual units and the rest of the system.''}, what we're finding is not the \emph{minimum} information partition but the \emph{modular} information partition. And if we seek the \emph{modular} information partition, there are several principled infotheoretic methods \cite{modules1, Rosvall_2008, Shalizi_2001,Olbrich_2010} for detecting causal modules in dynamic systems which run faster than IIT's current  $O(n!)$. Either way, it looks like there's options to save the IIT-community computation time, headache, or both.

\section{Conclusion}
IIT requires a measure of spacetime integration across a partition. Here we proposed four distinct measures of this derived from synergy. IIT's notion of integration can be a completely different thing from synergistic irreducibility---there's nothing inconsistent with that.  However, in our view, one of the greatest features of Integrated Information Theory is that \emph{even if} IIT is \emph{wrong} about consciousness, everything IIT remains scientifically and mathematically useful simply by replacing the word ``consciousness'' with ``irreducibility''. Anything that rigorizes this connection is worth taking seriously. Using synergy as a working operationalization of integration allows IIT to benefit from and contribute to a vastly larger pool of literature in information and complexity theory. Therefore, given no strident conflicts with IIT axioms and postulates, it seems sensible to at least consider synergistic irreducibility as a framework for spacetime integration. A final benefit is that the synergy framework naturally generalizes to the quantum and even algorithmic information theory \cite{Kolchinsky_2022} allowing for analogous extensions like \cite{Albantakis_2023}.

Possible downsides of the synergy approach are:
\begin{itemize}
\item No longer selects an individual cause-effect state $s^\prime = \{a^\prime, z^\prime\}$. Our understanding is that this structure arose from  IIT's ``principle of maximal existence''  which states ``when it comes to a requirement for existence, what exists is what exists the most.'' While this is a necessary postulate for defining Complexes, we believe that interpreting this postulate so literally across $A$ and $Z$ results in mathematical structures that preclude our strongest mathematical tools.

\item Synergy-based measures understand irreducibility differently. The comparison of system GET-GET in Figure \ref{fig:one} having maximum $\bphi$ but zero $\bpsi$ is the simplest example demonstrating this divergence. The $\phi$ measure has changed immensely between 2008--2025, and we see this change as surmountable. In measuring irreducibility, the fundamental question is ``irreducibility to what''? IIT's intuition is irreducibility beyond ``independent spatially disjoint parts''. Synergy formalizes and generalizes this to irreducibility beyond ``independent information streams'', but strongly holding the disjoint parts intuition can be accommodated by measure $\phi_s^{S3}$.

\item Although algorithms exist, the minimization in eq.~\eqref{eq:pcup} for $n \geq 4$ remains an area of research.
\end{itemize}

We view all four measures: $\phi_s^{S1}$, $\phi_s^{S2}$, $\phi_s^{S3}$, and $\phi_s^{S4}$, as viable candidates for IIT's future treatment of spacetime integration. Our assessment is that reframing integration as synergy improves one facet of IIT with negligible disruption to the rest of the theory and would be a net-positive in the continued evolution of IIT.

For those interested in irreducibility measures outside of the realm of IIT, all four measures work within standard Shannon theory, and you would replace all instances of $\mathcal{S}_{ID}$ with $\mathcal{S}$ and use eq.~\eqref{eq:mipbound} for the upperbound.

\subsection{Future Work}

\begin{itemize}
\item Although synergy for $n=2$ and $n=3$ are effectively solved, unfortunately $\phi_s^{S2}$ and $\phi_s^{S4}$ only become viable at $n=4$. Although there are approximations and tricks, greater research into the minimization of eq.~\eqref{eq:pcup} is needed to truly leverage those two measures.
\item The looseness of the upperbound in eq.~\eqref{eq:mipboundid} is painful and doesn't even change with merging predictors. We need to do better.
\end{itemize}

\subsection{Recommendations}

\begin{enumerate}
\item If we prefer to keep IIT4 as unchanged as possible, the best choice is $\phi_s^{S1}$ and normalize by eq.~\eqref{eq:mipboundid} when finding the MIP. Upsides: closest to the stated intuition of ``conjunction across time of spatial integrations''; minimally disruptive to IIT4. Downsides: Like $\phi_s^{2023}$, merely upperbounds the true conjunction.

\item If we prefer to adopt a measure needing as little additional mathematical research as possible, the best options are $\phi_s^{S1}$ and $\phi_s^{S3}$ limiting $\theta$ to bipartitions and tripartitions---this is very much solved. Upsides: Immediately have out-of-the-box practical measures. Downsides: If we're committed to the current practice in MIP normalization, we'll inevitably want to robustly estimate eq.~\eqref{eq:pcup} for arbitrary $n$. In the mean time, we could upperbound any of the four measures by leveraging that a union over terms is always greater than or equal to a maximum over those same terms. So we can bound any $\mathcal{S}_{ID}$-based measure by replacing the $p_\cup$ distribution in eq.~\eqref{eq:SID} with a max $x_i \in \{x_1, \ldots, x_n\}$.

\item If we prefer a time-symmetric measure closest to early intuitions of $\phi$ being irreducibility beyond ``independent disjoint parts'' \cite{iit2}, the choice is $\phi_s^{S3}$ as it is directly about irreducibility to spatially disjoint parts over time. Upsides: not spacetime-interchangeable; moderate; as a synergy of $k$ terms is faster to compute; handles past and future more elegantly than a minimum between past and future. Downsides: Changes the precise notion of integration.

\item If we prefer full spacetime interchangeability, the choices are $\phi_s^{S2}$ (loose) and $\phi_s^{S4}$ (strict). Upsides: spacetime-interchangeable. Downsides: Although $\phi_s^{S2} \lessgtr \phi_s^{2023}$, $\phi_s^{S2}$ might be felt too loose for IIT. Although befitting the principle of asking how much a system is ``unitary'', $\phi_s^{S4}$ might be felt too strict for IIT. Either would requires mathematical research to compute eq.~\eqref{eq:pcup} for at least $n = 4$.

\item If we insist that deterministic systems have zero integration, the choice is $\phi_s^{S4}$. Upsides: spacetime-interchangeable as well as captures in a strong sense how much the sources are ``unitary.'' Downsides: requires mathematical research to compute eq.~\eqref{eq:pcup} for at least $n = 4$..

\end{enumerate}

\textbf{Acknowledgements.} We used code from PyPhi \cite{PyPhi} and BROJA\_2PID \cite{makkeh2018broja}. We thank Artemy Kolchinsky, Will Mayner, Carlotta Langer, Abdullah Makkeh, and Christof Koch for their consultations.

\printbibliography

\clearpage

\input{appendix.tex}

\end{document}

%% file: appendix.tex
\part*{Appendix}
\appendix

\section{Another Way of Computing the Union Distribution}
\label{sect:otherway}

When computing the union information for an intrinsic difference, we chose the union distribution $p_\cup$ that matches the literature on synergistic mutual information. But this is not the only way to do it. Here's another way:

\begin{eqnarray}
\label{eq:pcup_otherway}
p_\cup^2\left( X_1, \ldots, X_n, y\right) &\equiv& \underset{p(\hat{X}_1,\ldots,\hat{X}_n,y)}{\operatorname{argmin} } \operatorname{ID}\left[ p(\hat{X}_1, \ldots, \hat{X}_n) \to y \right] \ \ \ \textnormal{s.t. $p(\hat{X}_i,y)=p(X_i,y)\ \forall i$} \nonumber \\
&=& \label{eq:pcup2min} \underset{p(\hat{X}_1,\ldots,\hat{X}_n,y)}{\operatorname{argmin} } \max_{x_1,\ldots,x_n} p(x_1, \ldots, x_n|y)  \log \frac{p(x_1, \ldots, x_n|y)}{p(x_1, \ldots, x_n)} \; . \\
& &  \textnormal{s.t. $p(\hat{X}_i,y)=p(X_i,y)\ \forall i$} \nonumber
\end{eqnarray}

The intuition in favor of $p_\cup^2$ has some appeal to it over eq.~\eqref{eq:pcup} as now all traces of $\operatorname{D_{KL}}$ have been removed. The problems with $p_\cup^2$ are:
\begin{enumerate}
    \item Foremost, ID is seemingly a non-convex function. And until proven it is, it is unclear how to robustly numerically find the minimum of eq.~\eqref{eq:pcup2min}.
    \item The mathematical object of $p_\cup$ in eq.~\eqref{eq:pcup} is reasonably well studied in the Partial Information Decomposition literature. Therefore, unless there's strong theoretical justification for preferring $p_\cup^2$, it's sensible to err on the side of choosing $p_\cup$.
\end{enumerate}

That said, if the IIT world prefers $p_\cup^2$ over $p_\cup$, further research into the minimization of eq.~\eqref{eq:pcup_otherway} could make it a viable path forward.

\clearpage
\section{Understanding the Doublet Networks}
\label{app:tutorial}
\begin{figure}[h!]
\centering
	\subfloat[ZERO-ZERO]{ \includegraphics[width=1.33in]{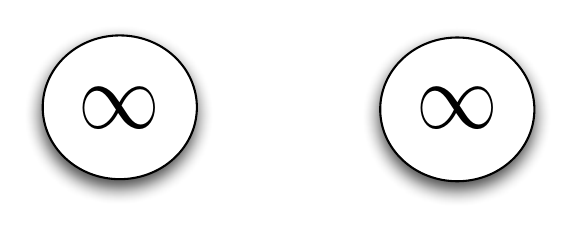} \label{fig:ZZ} }
	\subfloat[KEEP-ZERO]{ \includegraphics[width=1.33in]{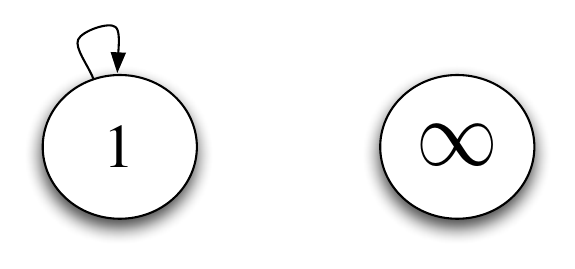} \label{fig:KZ} }
	\subfloat[GET-ZERO]{ \includegraphics[width=1.33in]{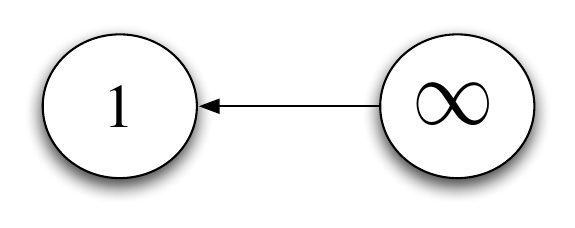} \label{fig:GZ} }
	\subfloat[KEEP-KEEP]{ \includegraphics[width=1.33in]{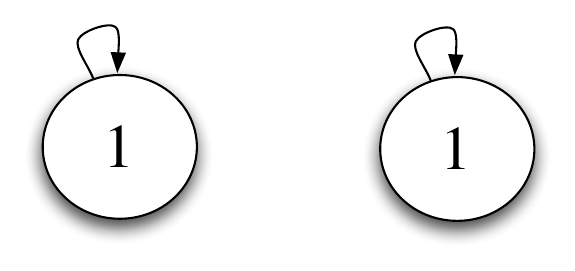} \label{fig:KK} }	
	
	\subfloat[GET-KEEP]{ \includegraphics[width=1.33in]{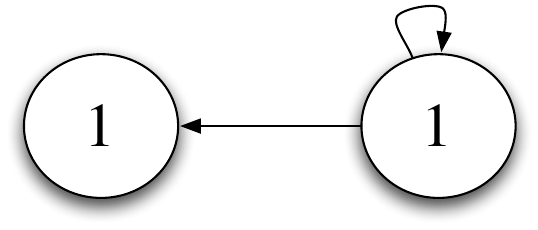} \label{fig:GK} }	
	\subfloat[GET-GET]{ \includegraphics[width=1.33in]{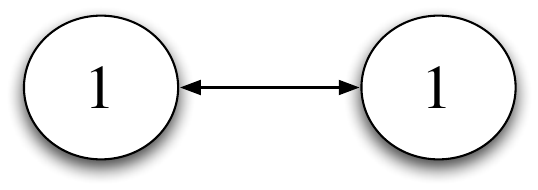} \label{fig:GG} }	
	\subfloat[AND-ZERO]{ \includegraphics[width=1.33in]{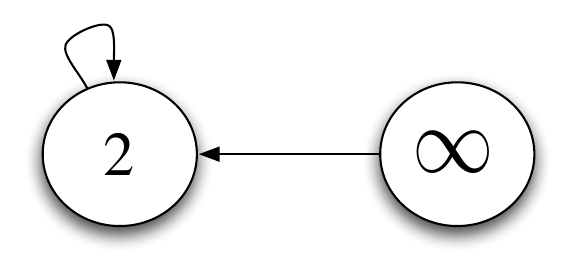} \label{fig:AZ_app} }		
	\subfloat[AND-XOR]{ \includegraphics[width=1.33in]{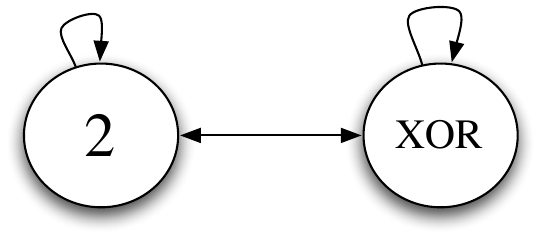} \label{fig:AX} }			

\vspace{0.2in}
\begin{tabular}{c c c c c c c c c c c} \toprule
	\multirow{2}{*}{$A$} & \ & ZERO- & KEEP- &GET- & KEEP- &GET-&GET-&AND-&AND-\\
	 & \ & ZERO& ZERO & ZERO & KEEP & KEEP & GET &ZERO & XOR \\
\midrule
	\bin{00} & $\to$ & \bin{00} & \bin{00} & \bin{00} & \bin{00} & \bin{00} & \bin{00} & \bin{00} & \bin{00} \\
	\bin{01} & $\to$ & \bin{00} & \bin{00} & \bin{10} & \bin{01} & \bin{11} & \bin{10} & \bin{00} & \bin{01} \\
	\bin{10} & $\to$ & \bin{00} & \bin{10} & \bin{00} & \bin{10} & \bin{00} & \bin{01} & \bin{00} & \bin{01} \\
	\bin{11} & $\to$ & \bin{00} & \bin{10} & \bin{10} & \bin{11} & \bin{11} & \bin{11} & \bin{10} & \bin{10} \\
	\bottomrule
\end{tabular}

\caption{Eight doublet networks with transition tables.}
\label{fig:tutorial}
\end{figure}

We present eight doublet networks and their transition tables so you can see how the network diagram specifies the transition table.  Figure \ref{fig:tutorial} shows eight network diagrams to build your intuition.  The number inside each node is that node's \emph{activation threshold}.  A node updates to \bin{1} (conceptually an ``ON'') if there at least as many of inputs ON as its activation threshold; eg, a node with an inscribed 2 updates to a \bin{1} if two or more incoming wires are ON.  An activation threshold of \scalebox{1.15}{$\infty$} means the node always updates to \bin{0} (conceptually an ``OFF'').  A binary string denotes the state of the network, read left to right.

We take the AND-ZERO network as an example.  Although the AND-ZERO network can never output \bin{01} or \bin{11}, we still consider states \bin{01}, \bin{11} as equally possible states in distribution $A$.

\clearpage

\section{Four Node Example}
\label{appendix:4node}

\begin{figure}[htb]
\centering

\begin{subfigure}{1.18in}
  \centering
  \includegraphics[width=\linewidth]{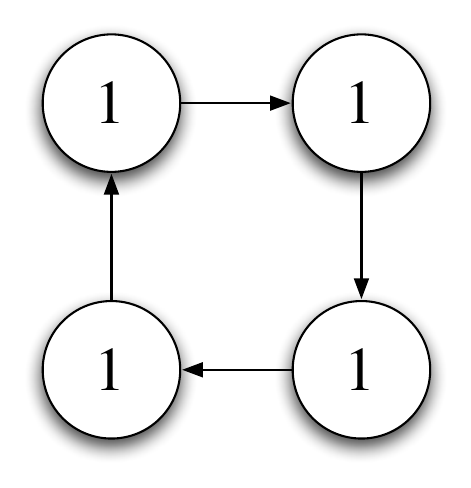}
  \caption{GET4}
  \label{fig:SHIFT4}
\end{subfigure}
\hfill
\begin{subfigure}{1.3in}
  \centering
  \includegraphics[width=\linewidth]{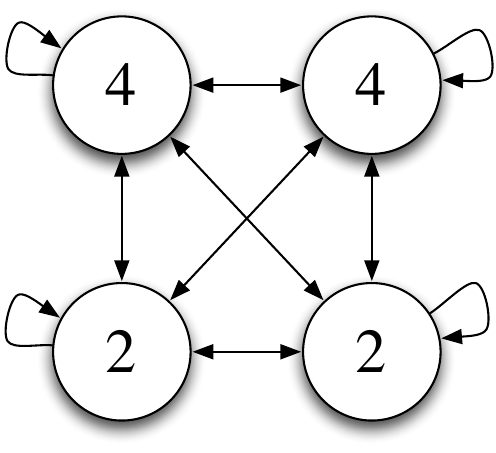}
  \caption{4422}
  \label{fig:4422}
\end{subfigure}
\hfill
\begin{subfigure}{1.3in}
  \centering
  \includegraphics[width=\linewidth]{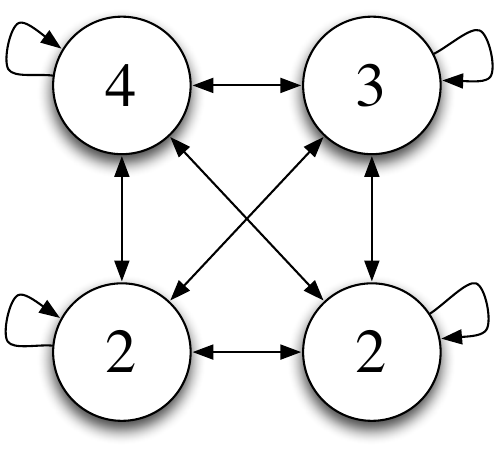}
  \caption{4322}
  \label{fig:4322}
\end{subfigure}
\hfill
\begin{subfigure}{1.3in}
  \centering
  \includegraphics[width=\linewidth]{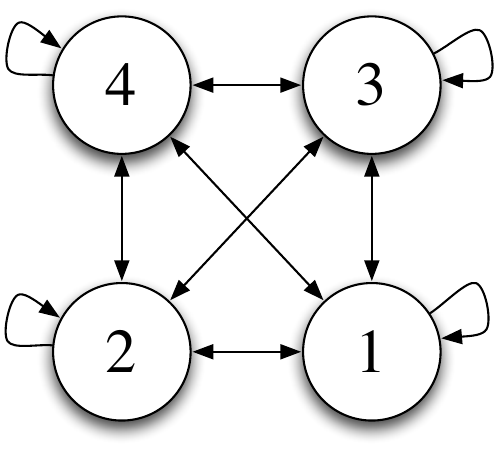}
  \caption{4321}
  \label{fig:4321}
\end{subfigure}

\vspace{0.3in}

\begin{tabular}{l c c c c c | c c} \toprule
	Network & $\operatorname{ID}\left[ A \to S \right]$ & $\displaystyle \min_{s} \phi^{2023}_c(s)$ & $\displaystyle \max_{s} \phi^{2023}_c(s)$ & $\bphi$ & $\bphics$ & $\Ip{A}{S}$ & $\mathcal{S}_c$ \\
\midrule
	GET4      & 4.000 & 2.000 & 2.000 & 2.000 & 0 & 4.000 & 0 \\
	4422       & 0.397 & 0 & 0 & 0 & 0.198 & 1.198 & 0.500 \\
	4322       & 0.568 & 0 & 0 & 0 & 0.369 & 1.805 & 0.875 \\
	4321       & 0.838 & 0 & 0 & 0 & 0.455 & 2.031 & 1.000 \\
\bottomrule
\end{tabular}
\caption{State-dependent $\phi^{2023}_c$ and $\bphi$ tell the same story---the $\phi^{2023}_c$ value of GET4 trounces the $\phi^{2023}_c$ of the other three networks. A more intuitive complexity measure would instead increase left to right, which $\bpsi$ does. To avoid claims of slight-of-hand, $\phi^{2023}_c$ is computed using a normalized MIP and $\langle \phi_s^{S1} \rangle$ is computed using an unnormalized MIP. We are seeing $\phi^{2023}_c$ at its best, and $\phi_c^{S1}$ at its worst. For readers preferring Shannon-based measures, we include the mutual information $\Ip{A}{S}$ and synergistic information $\mathcal{S}_c(\{A_1,A_2\} \to S)$.}
\label{fig:diverse4}
\end{figure}

\clearpage

\section{Comparing T-symmetric measures}
\label{app:comparefull}
This table has two purposes:
\begin{enumerate}

    \item \textbf{Comparing $\phi_s^{2023}$ vs $\phi_s^{S1}$.} As $\phi_s^{S1}$ was always zero, we didn't find this comparison pedagogically useful for understanding synergy. So in the paper we compared the forward-time versions: $\phi^{2008}$, $\phi^{2014}_c$, $\phi^{2023}_c$, $\phi^{2025}_c$, and $\phi_c^{S1}$.

    \item \textbf{Comparing prior time-symmetric IIT measures to $\phi_s^{S3}$.} After we understand synergy, how do the time-symmetric synergy measures $\phi_s^{S1}$ and $\phi_s^{S3}$ actually compare to the prior time-symmetric IIT measures?
\end{enumerate}

\begin{table}[h!]
\centering
\begin{tblr}{l|c c c c c c| c c c}
\toprule
\textbf{Doublet} & $\operatorname{ID}\left[ A \to S \right]$ & $\langle \phi^{2008}\rangle$ & $\langle \phi_s^{2014}\rangle$ & $\langle \phi_s^{2023}\rangle$ & $\langle \phi_s^{S1}\rangle$ & $\langle \phi_s^{S3}\rangle$ & $\Ip{A}{S}$ & $\mathcal{S}_s^{S1}$ & $\mathcal{S}_s^{S3}$ \\
\midrule
ZERO-ZERO & 0 & 0 & 0 & 0 & 0 & 0 & 0 & 0 & 0\\
KEEP-ZERO & 0.500 & 0 & 0 & 0 & 0 & 0 & 1.000 & 0 & 0\\
KEEP-KEEP & 2.000 & 0 & 0 & 0 & 0 & 0 & 2.000 & 0 & 0\\
GET-ZERO & 0.500 & 1.000 & 0 & 0 & 0 & 0 & 1.000 & 0 & 0\\
GET-KEEP & 0.500 & 0 & 0 & 0 & 0 & 0 & 1.000 & 0 & 0\\
GET-GET & 2.000 & 2.000 & 0 & 2.000 & 0 & 0 & 2.000 & 0 & 0\\
\midrule
AND-ZERO & 0.604 & 0.500 & 0 & 0 & 0 & 0.354 & 0.811 & 0 & 0.500\\
AND-KEEP & 1.250 & 0.189 & 0 & 0 & 0 & 0 & 1.500 & 0 & 0\\
AND-GET & 1.250 & 1.189 & 0.958 & 0.500 & 0 & 0 & 1.500 & 0 & 0\\
AND-AND & 0.604 & 0.189 & 0.500 & 0.500 & 0 & 0.354 & 0.811 & 0 & 0.500\\
AND-OR  & 1.250 & 0.877 & 1.623 & 0     & 0 & 0.125 & 1.500 & 0 & 0\\
AND-XOR & 1.250 & 1.189 & 0.458 & 0.500 & 0 & 0.500 & 1.500 & 0 & 0.500 \\
\midrule
XOR-ZERO & 0.500 & 1.000 & 0.500 & 0 & 0 & 0.250 & 1.000 & 0 & 0\\
XOR-KEEP & 2.000 & 1.000 & 0 & 0 & 0 & 0 & 2.000 & 0 & 0\\
XOR-GET & 2.000 & 2.000 & 1.500 & 2.000 & 0 & 0 & 2.000 & 0 & 0\\
XOR-AND & 1.250 & 1.189 & 0.458 & 0.500 & 0 & 0.500 & 1.500 & 0 & 0.500 \\
XOR-OR  & 1.250 & 1.189 & 1.665 & 0.500 & 0 & 0.125 & 1.500 & 0 & 0 \\
XOR-XOR & 0.500 & 1.000 & 1.000 & 1.000 & 0 & 0.500 & 1.000 & 0 & 1.000 \\
\midrule
GET3   & 3.000 & 2.000 & 2.000 & 2.000 & 0 & 0 & 3.000 & 0 & 0 \\
111     & 0.399 & 0.130 & 0.750 & 0.750 & 0 & 0.011 & 0.544 & 0 & 0 \\
121     & 0.677 & 0.941 & 0 & 0.052 & 0 & 0.299 & 1.406 & 0 & 0.500 \\
123     & 1.104 & 1.347 & 0.051 & 0  & 0 & 0.236 & 1.811 & 0 & 0 \\
\bottomrule
 \end{tblr}
 \caption{For completeness, we include the Shannon mutual information versions of both $\phi_s^{S1}$ and $\phi_s^{S3}$, denoted $\mathcal{S}_s^{S1}$ and $\mathcal{S}_s^{S3}$. Likewise, although $\phi^{2008}$ isn't T-symmetric, we include it.}
\end{table}

\clearpage